\documentclass[11pt]{article}

\usepackage[left=2cm, right=2cm, top=2.5cm, bottom=2.5cm]{geometry}
\usepackage[x11names]{xcolor}
\usepackage{fancyhdr, amssymb, cancel, amsmath, graphicx, pgfplots, tikz}

\newcommand{\stylecolor}{black}

\usepackage[colorlinks=true, urlcolor=orange, linkcolor=blue, citecolor=red, hyperindex=true, linktocpage=true]{hyperref}

\usepackage[explicit]{titlesec}

\newcommand*\sectionlabel{}
\titleformat{\section}
  {\gdef\sectionlabel{}
   \Large\bfseries\scshape}
  {\gdef\sectionlabel{\thesection. }}{0pt}
  {\begin{tikzpicture}[remember picture,overlay]
	\draw (-0.2, 0) node[right] {\textsf{\sectionlabel#1}};
	\draw[thick] (0, -0.4) -- (\textwidth, -0.4);
       \end{tikzpicture}
  }
\titlespacing*{\section}{0pt}{15pt}{20pt}

\newcommand*\subsectionlabel{}
\titleformat{\subsection}
  {\gdef\subsectionlabel{}
   \large\bfseries\scshape}
  {\gdef\subsectionlabel{\thesubsection.\ \  }}{0pt}
  {\begin{tikzpicture}[remember picture,overlay]
    	\draw (-0.15, 0) node[right] {\textsf{\subsectionlabel#1}};
       \end{tikzpicture}
  }
\titlespacing*{\subsection}{0pt}{10pt}{10pt}

\pgfplotsset{every axis legend/.append style={at={(1.02,1)},anchor=north west}}

\newcommand{\titletext}{Ising formulations of many NP problems}

\begin{document}

\pagestyle{fancy}
\renewcommand{\headrulewidth}{0pt}
\fancyhead{}

\fancyfoot{}
\fancyfoot[C] {\textsf{\textbf{\thepage}}}

\begin{equation*}
\begin{tikzpicture}
\draw (0.5\textwidth, -3) node[text width = \textwidth] {{\huge \begin{center} \color{\stylecolor} \textsf{\textbf{\titletext}} \end{center}}}; 
\end{tikzpicture}
\end{equation*}
\begin{equation*}
\begin{tikzpicture}
\draw (0.5\textwidth, 0.1) node[text width=\textwidth] {\large \color{black} \textsf{Andrew Lucas}};
\draw (0.5\textwidth, -0.5) node[text width=\textwidth] {\small \textsf{Department of Physics, Harvard University, Cambridge, MA, USA 02138}};
\end{tikzpicture}
\end{equation*}
\begin{equation*}
\begin{tikzpicture}
\draw (0.5\textwidth, -6) node[below, text width=0.8\textwidth] {\small We provide  Ising  formulations for many NP-complete and NP-hard problems, including all of Karp's 21 NP-complete problems.  This collects and extends  mappings to the Ising model from partitioning, covering and satisfiability.  In each case, the required number of spins is at most cubic in the size of the problem.   This work may be useful in designing adiabatic quantum optimization algorithms.};  
\end{tikzpicture}
\end{equation*}
\begin{equation*}
\begin{tikzpicture}
\draw (0, -13.1) node[right] {\texttt{lucas@fas.harvard.edu}};
\draw (\textwidth, -13.1) node[left] {\textsf{\today}};
\end{tikzpicture}
\end{equation*}

\tableofcontents

\section{Introduction}
\subsection{Quantum Adiabatic Optimization}
Recently, there has been much interest in the possibility of using  adiabatic quantum optimization (AQO) to solve NP-complete and NP-hard problems \cite{farhi, das}.\footnote{In this paper, when a generic statement is true for both NP-complete and NP-hard problems, we will refer to these problems as NP problems.  Formally this can be misleading as P is contained in NP, but for ease of notation we will simply write NP.}   This is due to the following trick:  suppose we have a quantum Hamiltonian $H_{\mathrm{P}}$ whose ground state encodes the solution to a problem of interest, and another Hamiltonian $H_0$, whose ground state is ``easy" (both to find and to prepare in an experimental setup).    Then, if we prepare a quantum system to be in the ground state of $H_0$, and then adiabatically change the Hamiltonian for a time $T$ according to \begin{equation}
H(t) = \left(1-\frac{t}{T}\right)H_0  + \frac{t}{T} H_{\mathrm{P}},
\end{equation}then if $T$ is large enough, and $H_0$ and $H_{\mathrm{P}}$ do not commute,  the quantum system will remain in the ground state for all times, by the adiabatic theorem of quantum mechanics.   At time $T$, measuring the quantum state will return a solution of our problem.

There has been debate about whether or not these algorithms would actually be useful: i.e., whether an adiabatic quantum optimizer would run any faster than classical algorithms \cite{altshuler, dickson2011, bapst, farhi2, farhi3, hen, jorg}, due to the fact that if the problem has size $N$, one typically finds \begin{equation}T= \mathrm{O}\left[ \exp\left(\alpha N^\beta\right)\right], \label{eq2}\end{equation}in order for the system to remain in the ground state, for positive coefficients $\alpha$ and $\beta$, as $N\rightarrow \infty$.   This is a consequence of the requirement that exponentially small energy gaps between the ground state of $H(t)$ and the first excited state, at some intermediate time, not lead to Landau-Zener transitions into excited states \cite{bapst}.\footnote{If one is only interested in approximate solutions (for example, finding a state whose energy per site is optimal, in the thermodynamic ($N\rightarrow\infty$) limit, as opposed to finding the exact ground state), one expects $T=\mathrm{O}(N^\gamma)$ \cite{bapst, santoro}.}      While it is unlikely that NP-complete problems can be solved in polynomial time by AQO, the coefficients $\alpha,\beta$ may be smaller than known classical algorithms, so there is still a possibility that an AQO algorithm may be more efficient than classical algorithms, on some classes of problems.

There has been substantial experimental progress towards building a device capable of running such algorithms \cite{sergio1, sergio2, dwave}, when the Hamiltonian $H_{\mathrm{P}}$ may be written as the quantum version of an Ising spin glass.  A classical Ising model can be written as a quadratic function of a set of $N$ spins $s_i=\pm 1$: \begin{equation}
H\left(s_1,\ldots, s_N\right) = -\sum_{i<j} J_{ij} s_i s_j - \sum_{i=1}^N h_i s_i.
\end{equation}The quantum version of this Hamiltonian is simply \begin{equation}
H_{\mathrm{P}} = H\left(\sigma_1^z,\ldots,\sigma^z_N\right)
\end{equation}where $\sigma^z_i$ is a Pauli matrix (a $2\times 2$ matrix, whose cousin $(1+\sigma_i^z)/2$ has eigenvectors $|0,1\rangle$ with eigenvalues $0,1$) acting on the $i^{\mathrm{th}}$ qubit in a Hilbert space of $N$ qubits $\lbrace |+\rangle, |-\rangle \rbrace^{\otimes N}$, and $J_{ij}$ and $h_i$ are real numbers.   We then choose $H_0$ to consist of transverse magnetic fields \cite{sergio1}:  \begin{equation}
H_0 =- h_0 \sum_{i=1}^N \sigma^x_i,
\end{equation}so that the ground state of $H_0$ is an equal superposition of all possible states in the eigenbasis of $H_{\mathrm{P}}$ (equivalent to the eigenbasis of the set of operators $\sigma^z_i$ ($i=1,\ldots,N$)).    This means that one does not expect any level crossings.\footnote{This is due to the fact that the eigenbases of $H_0$ and $H_{\mathrm{P}}$ are very different, and one has to tune (in our case) 2 parameters of a $2\times 2$ Hermitian matrix to find a matrix with degenerate eigenvalues (the identity matrix).  We only have one, $t$, and thus do not expect any degeneracies \cite{bapst}.}      For more work discussing the choice of $H$, see \cite{whitfield}.   Also, note that this class of Hamiltonians is not believed to be sufficient to build a universal adiabatic quantum computer \cite{biamonte2} -- at all times, $H(t)$ belongs to a special class of Hamiltonians called \emph{stoquastic} Hamiltonians \cite{bravyi}.

\subsection{Ising Spin Glasses}
 Ising spin glasses\footnote{In this paper, we will casually refer to the Ising models we are constructing as ``glasses", as they can be on general graphs and have  both positive and negative couplings $J_{ij}$.    There are various mathematical definitions for a spin glass,  none of which seem to capture properly the physical essence of a glass on all problems.   We will be liberal with our use of the word glass, and refer to any NP problem, formulated as an Ising model, as a glass.}  are known to be NP-hard problems for classical computers \cite{Barahona1982}, so it is natural to suspect intimate connections with all other NP  problems.  For the purposes of this paper, an NP-complete problem is always a decision problem with a yes or no answer (does the ground state of $H$ have energy $\le 0$?), whereas an NP-hard problem is an optimization problem (what is the ground state energy of $H$?).  The class of NP-complete problems includes a variety of notoriously hard problems, and has thus attracted much interest over the last 40 years \cite{karp, garey}.   Mathematically, because the decision form of the Ising model is NP-complete, there exists a polynomial time mapping to any other NP-complete problem.
 
  Analogies between the statistical physics of Ising spin glasses and NP problems have been frequently studied in the past \cite{fu1986, mezard1987, weigt}, and have been used to construct simulated annealing algorithms \cite{kirkpatrick} which have been quite fruitful in approximate algorithms for problems on classical computers.   These connections have suggested a \emph{physical} understanding of the emergence of hardness in these problems via a complex energy landscape with many local minima \cite{mezard}.  Conversely, computational hardness of solving glassy problems  has implications for the difficulty of the solutions to important scientific problems ranging from polymer folding \cite{bryngelson, berger} to memory \cite{hopfield} to collective decision making in economics and social sciences \cite{bouchaud, lucas2}.     Problems of practical scientific interest have already been encoded and solved (in simple instances) on experimental devices using Ising Hamiltonians \cite{xu, bian, perdomo, babbush2, neven2008, denchev}.
  
Finally, we note that Ising glasses often go by the name QUBO (quadratic unconstrained binary optimization), in the more mathematical literature \cite{boros, borosreview}.     Useful tricks have been developed to fix the values of some spins immediately \cite{rrr} and to decompose large QUBO problems into smaller ones \cite{billionnet}.
 
 \subsection{The Goal of This Paper}
 Mathematically, the fact that a problem is NP-complete means we can find a mapping to the decision form of the Ising model with a polynomial number of steps.   This mapping can be re-interpreted as a pseudo-Boolean optimization problem \cite{borosreview}.  As the constructions of these pseudo-Boolean optimization problems (or ``$p$-spin glasses") often lead to three-body or higher interactions in $H$ (e.g., terms of the form $s_1s_2s_3$), we then conclude by using ``gadgets" to reduce the problem to an Ising spin glass, by introducing a polynomial number of auxiliary spins which help to enforce the three-body interaction by multiple two-body interactions ($s_1s_2$) \cite{biamonte, babbush}.    As such, we can get from any NP-complete problem to the Hamiltonian of an Ising spin glass, whose decision problem (does the ground state have energy $\le 0$?) solves the NP-complete problem of interest.    Classical gadgets are useful for many problems in physics as the physical energy (Hamiltonian) contains three-body interactions, but they are also useful for writing down many algorithms in other fields (e.g. integer factorization \cite{peng}).
 
However, for generic problems, this is a very inefficient procedure, as the power of the polynomial can grow quite rapidly.    As such, the typical NP-complete problem (of size $N$) studied in the context of Ising glasses is very straightforward to write as a glass with $N$ spins (such as number partitioning, or satisfiability).     The primary purpose of this paper is to present constructions of Ising Hamiltonians for problems where finding a choice of Hamiltonian is a bit subtle; for pedagogical purposes, we will also provide a review of some of the simple maps from partitioning and satisfiability to an Ising spin glass.     In particular, we will describe how ``all of the famous NP problems"\footnote{No offense to anyone whose problems have been left out.}  \cite{karp, garey} can be written down as Ising models with a polynomial number of spins which scales no faster than $N^3$.   For most of this paper, we will find it no more difficult to solve the NP-hard optimization problem vs. the NP-complete decision problem, and as such we will usually focus on the optimization problems.   The techniques employed in this paper, which are rare elsewhere in the quantum computation literature, are primarily of a few flavors, which roughly correspond to the tackling the following issues:  minimax optimization problems, problems with inequalities as constraints (for example, $n\ge 1$, as opposed to $n=1$), and problems which ask global questions about graphs.     The methods we use to phrase these problems as Ising glasses generalize very naturally.   

\subsection{What Problems Are Easy (to Embed) on Experimental AQO Devices?}
We hope that the reader may be inspired, after reading this paper, to think about solving some of these classical computing problems, or others like them, on experimental devices implementing AQO.   Towards this end, the reader should look for three things in the implementations in this paper.   The first is the number of spins required to encode the problem.   In some instances, the ``logical spins/bits" (the spins which are required to encode a solution of the problem) are the only spins required; but in general, we may require auxiliary ``ancilla spins/bits", which are required to enforce constraints in the problem.   Sometimes, the number of ancilla bits required can be quite large, and can be the dominant fraction of the spins in the Hamiltonian.    Another thing to watch out for is the possibility that large separations of energy scales are required:  e.g., the ratio of couplings $J_{12}/J_{23}$ in some Ising glass is proportional to $N$, the size of the problem being studied.    A final thing to note is whether or not the graph must be highly connected:  does the typical degree of vertices on the \emph{Ising embedding graph} (not the graph associated with the NP problem) scale linearly with $N$?

It is probably evident why we do not want too many ancilla bits -- this simply means we can only encode smaller problems on the same size piece of hardware.   It is a bit more subtle to understand  why complete graphs, or separations of energy scales, are problematic.   It is probable that the successful experimental implementations of AQO with the most qubits are on devices generated by DWave Systems \cite{sergio1, sergio2, dwave}.\footnote{These devices use quantum annealing, which is the finite temperature generalization of AQO.   For this paper, this is not an important issue, although it can certainly be relevant to experiments.}    As such, we now discuss the ease with which these Hamiltonians can be encoded onto such a device.    These devices may only encode problems via a ``chimera" graph.   The primary problem with Hamiltonians on a complete graph is that it is inefficient \cite{choiem1, choiem2} to embed complete graphs onto the chimera graph.     A primary difficulty is demonstrated by the following simple case:  a node $v$ in the complete graph must be mapped two a pair of nodes $u$ and $w$ on the chimera graph, with the coupling $J_{uw}$ large compared to other scales in the problem, to ensure that $s_u=s_w$ (so these nodes effectively act as one spin).    A second problem is that some of the Hamiltonians require separations of energy scales.   However, in practice, these devices may only encode couplings constants of $1, \ldots, 16$, due to experimental uncertainties \cite{sergio1, sergio2, dwave}.   This means that it is unlikely that, for very connected graphs, one may successfully encode any $H$ with a separation of energy scales.    A final challenge is that sometimes couplings or qubits are broken -- at this early stage in the hardware development, optimal algorithms have embeddings which are insensitive to this possibility \cite{klymko}.


\section{Partitioning Problems} \label{sec2}
The first class of problems we will study are partitioning problems, which (as the name suggests) are problems about dividing a set into two subsets.     These maps are celebrated in the spin glass community \cite{mezard}, as they helped physicists realize the possibility of using spin glass technology to understand computational hardness in random ensembles of computing problems.  For completeness, we review these mappings here, and present a new one based on similar ideas (the clique problem).
\subsection{Number Partitioning}
Number partitioning asks the following:   given a set of $N$ positive numbers $S = \lbrace n_1,\ldots, n_N\rbrace$, is there a partition of this set of numbers into two disjoint subsets $R$ and $S-R$, such that the sum of the elements in both sets is the same?   For example, can one divide a set of assets with values $n_1,\ldots,n_N$, fairly between two people?  This problem is known to be NP-complete \cite{karp}.    This can be phrased trivially as an Ising model as follows.   Let $n_i$ ($i=1,\ldots, N=|S|$) describe the numbers in set $S$, and let \begin{equation}
H= A\left(\sum_{i=1}^N n_is_i\right)^2
\end{equation}be an energy function, where $s_i=\pm 1$ is an Ising spin variable.     Here $A>0$ is some positive constant.   Typically, such constants are scaled to 1 in the literature, but for simplicity we will retain them, since in many formulations a separation of energy scales will prove useful,  and retaining each scale can make it easier to follow conceptually.   Classical studies of this problem are slightly easier if the square above is replaced with absolute value \cite{mezard}.

  It is clear that if there is a solution to the Ising model with $H=0$, then there is a configuration of spins where the sum of the $n_i$ for the $+1$ spins is the same for the sum of the $n_i$ for the $-1$ spins.    Thus, if the ground state energy is $H=0$, there is a solution to the number partitioning problem.     
  
  This Ising glass has \emph{degeneracies} -- i.e., there are always at least two different solutions to the problem.   This can be seen by noting that if $s_i^*$ denotes a solution to the problem, then $-s_i^*$ is also a solution.  Physically, this corresponds to the fact that we do not care which set is labeled as $\pm$.   In the spin glass literature, the change $s_i \rightarrow -s_i$, which does not change the form of $H$, is often (rather loosely) called a \emph{gauge transformation}.   The existence of a gauge transformation which leaves the couplings unchanged (as there are no linear terms) implies that all energy levels of $H$ are degenerate.   It is possible that there are $2m$ ground states (with $m>1$).   This means that there are $m$ physically distinct solutions to the computational problem.   We only need to find one of them to be happy with our adiabatic quantum algorithm.   We can remove this double degeneracy by fixing $s_1=1$.   This also allows us to remove one spin:  now only $s_2,\ldots,s_N$ are included on the graph, and $s_1$ serves as an effective magnetic field.   So in general, we require $N-1$ spins, which live on a complete graph, to encode this problem.
  
If the ground state has $H>0$, we know that there are no solutions to the partitioning problem, but the ground state we do find is (one of) the best possible solutions, in the sense that it minimizes the mismatch.   Minimizing this mismatch is an NP-hard problem, and we see that we do not require any more fancy footwork to solve the optimization problem -- the same Hamiltonian does the trick.
\subsection{Graph Partitioning}
Graph partitioning is the original \cite{fu1986} example of a map between the physics of Ising spin glasses and NP-complete problems.     Let us consider an undirected graph $G=(V,E)$.  with an even number $N=|V|$ of vertices.   We ask:  what is a partition of the set $V$ into two subsets of equal size $N/2$ such that the number of edges connecting the two subsets is minimized?   This problem has many applications:  finding these partitions can allow us to run some graph algorithms in parallel on the two partitions, and then make some modifications due to the few connecting edges at the end \cite{billionnet}.      Graph partitioning is known to be an NP-hard problem; the corresponding decision problem (are there less than $k$ edges conecting the two sets?) is NP-complete \cite{karp}.   We will place an Ising spin $s_v=\pm 1$ on each vertex $v\in V$ on the graph, and we will let $+1$ and $-1$ denote the vertex being in either the $+$ set or the $-$ set.  We solve this with an energy functional consisting of two components: \begin{equation}
H=H_A+H_B
\end{equation}where \begin{equation}
H_A = A\left(\sum_{n=1}^N s_i\right)^2
\end{equation}is an energy which provides a penalty if the number of elements in the + set is not equal to the number in the $-$ set, and \begin{equation}
H_B = B\sum_{(uv) \in E} \frac{1-s_us_v}{2}
\end{equation}is a term which provides an energy penalty $B$ for each time that an edge connects vertices from different subsets.      If $B>0$, then we wish to minimize the number of edges between the two subsets;  if $B<0$, we will choose to maximize this number.    Should we choose $B<0$, we must ensure that $B$ is small enough so that it is never favorable to violate the constraint of $H_A$ in order to minimize energy.    To determine a rather simple lower bound on $A$, we ask the question:  what is the minimum value of $\Delta H_B$ -- the change in the energy contributed by the $B$-term -- if we violate the $A$ constraint once.   It is easy to see that the penalty for violating the $A$-constraint is $\Delta H_A \ge 4A$.    The best gain we can get by flipping a spin is to  gain an energy of $B\min(\Delta,N/2)$, where $\Delta$ is the maximal degree of $G$.\footnote{The reason we can use $N/2$ in this formula instead of $N$ has to do with the fact that we are ``perturbing" a solution where $H_A=0$.  Due to the fact that the $H_A$ constraint is very penalizing if it is violated by having many spins in the same partition, it is easy to see that cases where an energy gain of $(N-1)B$ can be obtained by flipping a spin are very energetically penalized, and not relevant to the discussion.}   We conclude \begin{equation}
\frac{A}{B} \ge \frac{\min(2\Delta, N)}{8}.
\end{equation}$N$ spins on a complete graph are required to encode this problem.

This Hamiltonian is invariant under the same gauge transformation $s_i\rightarrow -s_i$.   We conclude that we can always remove one spin by fixing a single vertex to be in the $+$ set.

We have written $H$ in a slightly different form than the original \cite{fu1986}, which employed a constraint on the space of solutions to the problem, that \begin{equation}
\sum_{i=1}^N s_i = 0.
\end{equation}   We will want none of our formulations to do this (i.e., we wish to solve the unconstrained optimization problem), as the experimental hardware that is being built for quantum optimization can only solve unconstrained problems.   Instead, we encode constraint equations by making penalty Hamiltonians which raise the energy of a state which violates them.
\subsection{Cliques}
A clique of size $K$ in an undirected graph $G=(V,E)$ is a subset $W\subseteq V$ of the vertices, of size $|W|=K$, such that the subgraph $(W,E_W)$   (where $E_W$ is the edge set $E$ restricted to edges between nodes in $W$) is a complete graph -- i.e., all possible $K(K-1)/2$ edges in the graph are present, because every vertex in the clique has an edge to every other vertex in the clique.   Cliques in social networks can be useful as they are ``communities of friends"; finding anomalously large cliques is also a key sign that there is structure in a graph which may appear to otherwise be random \cite{alon}.   The NP-complete decision problem of whether or not a clique of size $K$ exists \cite{karp} can be written as an Ising-like model, as follows.  We place a spin variable $s_v=\pm 1$ on each vertex $v\in V$ of the graph.    In general, in this paper, for a spin variable $s_\alpha$, we will define the binary bit variable \begin{equation}
x_\alpha \equiv \frac{s_\alpha+1}{2}.
\end{equation}  It will typically be more convenient to phrase the energies in terms of this variable $x_\alpha$, as it will be for this problem.   Note that any energy functional which was quadratic in $s_v$ will remain quadratic in $x_v$, and vice versa, so we are free to use either variable.   We then choose \begin{equation}
H= A\left(K-\sum_v x_v\right)^2 + B\left[\frac{K(K-1)}{2}-\sum_{(uv)\in E} x_ux_v\right]
\end{equation}where $A,B>0$ are positive constants.    We want the ground state of this Hamiltonian is $H=0$ if and only if a clique of size $K$ exists.   It is easy to see that $H=0$ if there is a clique of size $K$.   However, we wish to now show that $H\ne 0$ for any other solution.   It is easy to see that if there are $n$ $x_v$s which are 1, that the minimum possible value of $H$ is \begin{equation}
H_{\mathrm{min}}(n) = A(n-K)^2 + B\frac{K(K-1) - n(n-1)}{2} = (n-K)\left[A(n-K) - B\frac{n+K-1}{2}\right].
\end{equation}The most ``dangerous" possible value of $n=1+K$.   We can easily see that so long as $A>KB$, $H_{\mathrm{min}}(K+1)>0$.   We finally note that, given a ground state solution, it is of course easy to read off from the $x_v$ which $K$ nodes form a clique.    $N$ spins on a complete graph are required to solve this problem.

A quantum algorithm for this NP-complete problem can be made slightly more efficient so long as the initial state can be carefully prepared \cite{childs}.

The NP-hard version of the clique problem asks us to find (one of) the \emph{largest} cliques in a graph.   We can modify the above Hamiltonian to account for this, by adding an extra variable $y_i$ ($i=2,\ldots, \Delta$), which is 1 if the largest clique has size $i$, and 0 otherwise.   Let $H= H_A+H_B+H_C$ where \begin{equation}
H_A = A \left(1-\sum_{i=2}^N y_i\right)^2 + A\left(\sum_{i=2}^n ny_n - \sum_v x_v\right)^2
\end{equation}and \begin{equation}
H_B = B\left[\frac{1}{2}\left(\sum_{i=2}^N ny_n\right)\left(-1+\sum_{i=2}^N ny_n\right) - \sum_{(uv)\in E} x_u x_v\right].
\end{equation}We want cliques to satisfy $H_A=H_B=0$, and to be the only ground states.   The Hamiltonian above satisfies this if $A/B$ is large enough so the constraints of $H_A=0$ are always satisfied -- we can see this by noting that the first term of $H_A$ forces us to pick only one value of $y_i=n$, and the second term fixes us to choose $n$ vertices.   Then $H_B=0$ ensures that we have a clique.  Similarly to the discussion above, we see that the absence of negative energy states requires $A>NB$.   If the maximal degree  of the graph is $\Delta$, this can be simplified to $A>\Delta B$.    Now that we know that all  ground states are cliques,\footnote{The ground state has $H=0$ so long as the edge set is non-empty:  any connected pair of edges is a clique of size 2.} we have to find the state with the smallest value of $y_n$.   This can be obtained by choosing \begin{equation}
H = -C\sum_v x_v,
\end{equation}where $C>0$ is some constant.  If $C$ is small enough, then the ground state energy is $H=-CK$, where $K$ is the size of the largest clique in the graph.   To determine an upper bound on $C$, so that we solve the cliques problem (as opposed to some other problem), we need to make sure that it is never favorable to color an extra vertex, at the expense of mildly violating the $H_A$ constraint.   The penalty for coloring one extra vertex, given $y_i=n$, is at minimum $A - nB - C$.   We conclude that we must choose \begin{equation}
C<A-nB.
\end{equation}So, for example, we could take $A=(\Delta+2)B$, and $B=C$.    

\subsection{Reducing $N$ to $\log N$ Spins in Some Constraints}
There is a trick which can be used to dramatically reduce the number of extra $y_i$ spins which must be added, in the NP-hard version of the clique problem above \cite{schrijver}.   In general, this trick is usable throughout this paper, as we will see similar constructions of auxiliary $y$ bits appearing repeatedly.   

We know that we want to encode a variable which can take the values $2,\ldots, N$ (or $\Delta$, if we know the maximal degree of the graph -- the argument is identical either way).   For simplicity, suppose we wish to encode the values $1,\ldots, N$ (this is a negligible difference, in the large $N$ limit).    Define an integer $M$ so that \begin{equation}
2^{M}\le N < 2^{M+1}.
\end{equation} Alternatively, $M = \lfloor \log N\rfloor$ -- in this paper, the base 2 is implied in the logarithm.   In this case, we only need $M+1$ binary variables: $y_0,\ldots, y_M$, instead of $N$ binary variables, $y_1,\ldots, y_N$, to encode a variable which can take $N$ values.   It is easy to check that replacing \begin{equation}
\sum_{n=1}^N ny_n  \rightarrow \sum_{n=0}^{M-1} 2^n y_n + \left(N+1-2^M\right)y_M
\end{equation}solves the same clique problem, without loss of generality.   (This is true in general for all of our NP problems.)   If $N\ne 2^{M+1}-1$, the ground state may be degenerate, as the summation of $y$s to a given integer is not always unique.     When actually encoding these problems for computational purposes, of course, this trick should be used, but for pedagogy and simplicity we will not write it out explicitly for the remainder of the paper.

Using this trick, we note that solving the NP-hard version of the cliques problem requires $N+1+\lfloor \log\Delta \rfloor$ spins.

\section{Binary Integer Linear Programming}
Let $x_1,\ldots, x_N$ be $N$ binary variables, which we arrange into a vector $\mathbf{x}$.   The binary integer linear programming (ILP) problem asks:   what is the largest value of $\mathbf{c}\cdot \mathbf{x}$, for some vector $\mathbf{c}$, given a constraint \begin{equation}
\mathsf{S}\mathbf{x} = \mathbf{b}  \label{axb}
\end{equation}with $\mathsf{S}$ an $m\times N$ matrix and $\mathbf{b}$ a vector with $m$ components.     This is NP-hard \cite{karp}, with a corresponding NP-complete decision problem.   Many problems can be posed as ILP: e.g., a supplier who wants to maximize profit, given regulatory constraints  \cite{schrijver}. 

The Ising Hamiltonian corresponding to this problem can be constructed as follows.  Let $H=H_A+H_B$ where \begin{equation}
H_A = A\sum_{j=1}^m \left[b_j - \sum_{i=1}^N S_{ji}x_i\right]^2  
\end{equation}and $A>0$ is a constant.   The ground states of $H_A=0$ enforce (if such a ground state exists, of course!) the constraint that $\mathsf{S}\mathbf{x}=\mathbf{b}$.     Then we set \begin{equation}
H_B = -B\sum_{i=1}^N c_ix_i.
\end{equation}with $B\ll A$ another positive constant.  

To find constraints on the required ratio $A/B$, we proceed similarly to before.   For simplicity, let us assume that the constraint Eq. (\ref{axb}) can be satisfied for some choice of $\mathbf{x}$.    For such a choice, the largest possible value of $-\Delta H_B$ is, in principle, $B\mathcal{C}$, where\begin{equation}
\mathcal{C} = \sum_{i=1}^N \max(c_i,0).
\end{equation}The smallest possible value of $\Delta H_A$ is related to the properties of the matrix $\mathsf{S}$, and would occur if we only violate a single constraint, and violate that constraint by the smallest possible amount, given by \begin{equation}
\mathcal{S} \equiv \min_{\sigma_i \in \lbrace0,1\rbrace,\; j }\left(\max\left[1,\frac{1}{2}\sum_{i}(-1)^{\sigma_i}S_{ji}\right]\right).
\end{equation}This bound could be made better if we know more specific properties of $\mathsf{S}$ and/or $\mathbf{b}$.   We conclude \begin{equation}
\frac{A}{B} \ge \frac{\mathcal{C}}{\mathcal{S}}.
\end{equation}If the coefficients $c_i$ and $S_{ij}$ are O(1) integers, we have $\mathcal{C} \le N\max(c_i)$, and $\mathcal{S}\ge 1$, so we conclude $A/B \gtrsim N$.

\section{Covering and Packing Problems}\label{sec3}
In this section, we discuss another simple class of mappings from NP problems to Ising models: ``covering" and ``packing" problems.  These problems can often be thought of as asking: how can I pick elements out of a set (such as vertices out of a graph's vertex set) so that they ``cover" the graph in some simple way (e.g., removing them makes the edge set empty).   In this class of problems, there are constraints which must be exactly satisfied.    Many of the problems described below are often discussed in the literature, but again we review them here for completeness.   We conclude the section with the minimal maximal matching problem, which is a slightly more involved problem that has not been discussed in the AQO literature before.

These are, by far, the most popular class of problems discussed in the AQO literature.   As we mentioned in the introduction, this is because this is the \emph{only} class of NP problems (discussed in this paper) for which it is easy to embed the problem via a graph which is not complete (or close to complete). 

\subsection{Exact Cover}
The exact cover problem goes as follows:    consider a set $U = \lbrace 1,\ldots, n\rbrace$, and subsets $V_i \subseteq U$ $(i=1,\ldots, N$) such that \begin{equation}
U = \bigcup_i V_i.
\end{equation}The question is:   is there a subset of the set of sets $\lbrace V_i\rbrace$, called $R$, such that the elements of $R$ are disjoint sets, and the union of the elements of $R$ is $U$?      This problem was described in \cite{choi2010} but for simplicity, we repeat it here.   This decision problem is NP-complete \cite{karp}.  The Hamiltonian we use is\begin{equation}
H_A = A\sum_{\alpha=1}^n \left(1-\sum_{i:\alpha\in V_i} x_i\right)^2.
\end{equation}In the above Hamiltonian $\alpha$ denotes the elements of $U$, while $i$ denotes the subsets $V_i$.   $H_A=0$ precisely when every element is included exactly one time, which implies that the unions are disjoint.   The existence of a ground state of energy $H=0$ corresponds to the existence of a solution to the exact cover problem.   If this ground state is degenerate, there are multiple solutions.   $N$ spins are required.

It is also straightforward to extend this, and find the \emph{smallest} exact cover (this makes the problem NP-hard).   This is done by simply adding a second energy scale: $H=H_A+H_B$, with $H_A$ given above, and \begin{equation}
H_B = B\sum_i x_i.
\end{equation}The ground state of this model will be $mB$, where $m$ is the smallest number of subsets required.     To find the ratio $A/B$ required to encode the correct problem, we note that the worst case scenario is that there are a very small number of subsets with a single common element, whose union is $U$.   To ensure this does not happen, one can set $A>nB$.\footnote{The example where $V=\lbrace\lbrace 1,2\rbrace,\lbrace3\rbrace,\ldots,\lbrace n\rbrace, \lbrace 2,\ldots,n\rbrace\rbrace$ shows that to leading order in $n$, this bound is optimal.}


\subsection{Set Packing}
Let us consider the same setup as the previous problem, but now ask a different question:  what is the largest number of subsets $V_i$ which are all disjoint?   This is called the set packing problem; this optimization problem is NP-hard \cite{karp}.    To do this, we use $H=H_A+H_B$: \begin{equation}
H_A=A\sum_{i,j:V_i\cap V_j\neq \emptyset} x_i x_j,
\end{equation}which is minimized only when all subsets are disjoint.    Then, we use \begin{equation}
H_B = -B\sum_i x_i
\end{equation}which simply counts the number of sets we included.     Choosing $B<A$ ensures that it is never favorable to violate the constraint $H_A$ (since there will always be a penalty of at least $A$ per extra set included) \cite{dickson2011}.

Note that an isomorphic formulation of this problem, in the context of graph theory is as follows:   let us consider the sets to be encoded in an undirected graph $G=(V,E)$, where each set $V_i$ maps to to a vertex $i\in V$.    An edge $ij\in E$ exists when $V_i\cap V_j$ is nonempty.  It is straightforward to see that if we replace \begin{equation}
H_A = A\sum_{ij\in E} x_ix_j 
\end{equation}that the question of what is the maximal number of vertices which may be ``colored"  ($x_i=1$) such that no two colored vertices are connected by an edge, is exactly equivalent to the set packing problem described above.   This version is called the maximal independent set (MIS) problem.

\subsection{Vertex Cover}
Given an undirected graph $G=(V,E)$, what is the smallest number of vertices that can be ``colored" such that every edge is incident to a colored vertex?   This is NP-hard; the decision form is NP-complete \cite{karp}.   Let $x_v$ be a binary variable on each vertex, which is 1 if it is colored, and 0 if it is not colored.    Our Hamiltonian will be $H=H_A+H_B$.   The constraint that every edge has at least colored vertex is encoded in $H_A$:\begin{equation}
H_A = A\sum_{uv\in E}(1-x_u)(1-x_v).
\end{equation}Then, we want to minimize the number of colored vertices with $H_B$: \begin{equation}
H_B = B\sum_v x_v
\end{equation}
Choose $B<A$, as if we uncolor any vertex that ruins the solution, at least one edge will no longer connect to a colored vertex.    The number of spins required is $|V|$, the size of the vertex set.

\subsection{Satisfiability}
Satisfiability is one of the most famous NP-complete problems \cite{karp}.   Every satisfiability problem can be written as a so-called 3SAT problem in conjunctive normal form (and this algorithm takes only polynomial steps/time) and so we will focus for simplicity on this case.    In this case, we ask whether \begin{equation}
\Psi = C_1\wedge C_2\cdots \wedge C_m
\end{equation}can take on the value of true -- i.e., every $C_i$ for $1\le i\le m$ is true, where the form of each $C_i$ is: \begin{equation}
C_i = y_{i_1}\vee y_{i_2}\vee y_{i_3}
\end{equation}Here $y_{i_1}$, $y_{i_2}$ and $y_{i_3}$ are selected from another set of Boolean variables: $x_1,\ldots, x_N, \overline{x}_1,\ldots, \overline{x}_N$.   This is a very brief description of satisfiability; physicists who are unfamiliar with this problem should read appropriate chapters of \cite{mezard}.

There is a well-known reduction of 3SAT to MIS \cite{choi2010} which we reproduce here, for completeness.    Consider solving the set packing problem on a graph $G$ with $3m$ nodes, which we construct as follows.    For each clause $C_i$, we add 3 nodes to the graph, and connect each node to the other 3.   After this step, if there is a $y_1$ and $y_2$ such that $y_1=\overline{y}_2$, then we also add an edge between these two nodes.     Solving MIS on this graph, and asking whether the solution has exactly $m$ nodes, is equivalent to solving the 3SAT problem.  This can be seen as follows:   if a solution to the 3SAT problem exists, only one element of each clause needs to be true -- if more are true, that is also acceptable, but we must have that one is true, so let us choose to color the vertex corresponding to the variable which is true.   However, we may also not choose to have both $x_1$ be true and $\overline{x}_1$ be true, so we are required to connect all such points with an edge.   Since the graph is made up of $m$ connected triangles, the only way to color $m$ vertices if each vertex is in a distinct triangle, so there must be an element of each clause $C_i$ which is true.

Note that we can solve an NP-hard version of this problem (if we have to violate some clauses, what is the fewest number?), by solving the optimization version of the MIS problem.

\subsection{Minimal Maximal Matching}

The minimal maximal (minimax) matching problem on a graph is defined as follows:   let $G=(V,E)$ denote an undirected graph, and let $C\subseteq E$ be a proposed ``coloring".    The  constraints on $C$ are as follows:   for each edge in $C$, let us color the two vertices it connects:  i.e. let $D = \bigcup_{e\in C} \partial e$.    We will then demand that:  no two edges in $C$ share a vertex (if $e_1,e_2 \in C$, $\partial e_1 \cap \partial e_2 = \emptyset$) and that if $u,v\in D$, that $(uv) \notin E$.   This is NP-hard; the decision problem is NP-complete \cite{garey}.  This is minimal in that we cannot add any more edges to $C$ (coloring any appropriate vertices) without violating the first constraint, and maximal in the sense that the trivial empty set solution is not allowed -- we must include all edges between uncolored vertices.   

Note that, from this point on in this paper, we have not found any of the Ising formulations of this paper in the literature.   

We will use the spins on the graph to model whether or not an edge is colored.   Let us use the binary variable $x_e$ to denote whether or not an edge is colored;  thus, the number of spins is $|E| = \mathrm{O}(\Delta N)$, the size of the edge set; as before, $\Delta$ represents the maximal degree.  To encode this problem, we use a series of three Hamiltonians:\begin{equation}
H=H_A+H_B+H_C.
\end{equation}The first and largest term, $H_A$, will impose the constraint that no vertex has two colored edges.   This can be done by setting \begin{equation}
H_A = A\sum_v \sum_{\lbrace e_1,e_2\rbrace\subset \partial v} x_{e_1}x_{e_2}.
\end{equation}Here $A>0$ is a positive energy, and $\partial v$ corresponds to the subset of $E$ of edges which connect to $v$.   Thus the ground states consist of $H_A=0$;  if $H_A>0$, it is because there is a vertex where two of its edges are colored.   

We also can define, \emph{for states with $H_A=0$}, the variable \begin{equation}
y_v \equiv \left\lbrace \begin{array}{ll} 1 &\  v\text{ has a colored edge} \\ 0 &\  v\text{ has no colored edges}  \end{array} \right.= \sum_{e\in\partial v} x_e.
\end{equation}We stress that this definition is only valid for states with $H_A=0$, since in these states each vertex has either 0 or 1 colored edges.    We then define the energy $H_B$, such that solutions to the minimax coloring problem also have $H_B=0$.   Since we have already constrained the number of colored edges per vertex, we choose $H_B$ to raise the energy of all solutions where there exists a possible edge which can be colored, yet still not violate the coloring condition, out of the ground state.   To do this, we can sum over all edges in the graph, and check whether or not the edge connects two vertices, neither of which are colored: \begin{equation}
H_B = B\sum_{e=(uv)} (1-y_u)(1-y_v).
\end{equation}
 Note that since, $1-y_v$ can be negative, we must choose $B>0$ to be small enough.   To bound $B$, we note that the only problem (a negative term in $H_B$) comes when $y_u=0$, $y_v>1$, and $(uv)\in E$.   Suppose that $m$ of $v$'s neighbors have $y_u=0$.   Then, the contributions to $H_A$ and $H_B$ associated to node $v$ are given by \begin{equation}
 H_v = A\frac{y_v(y_v-1)}{2}  - B(y_u-1)m.
 \end{equation}Note that $m+y_u \le k$, if $k$ is the degree of node $v$.   Putting all of this together, we conclude that if $\Delta$ is the maximal degree in the graph, because the worst case scenario is $y_v=2$, $m=\Delta-2$, if we pick \begin{equation}
 A>(\Delta-2)B,
 \end{equation}then it is never favorable to have any $y_v>1$.   This will ensure that a ground state of $H_A+H_B$ will have   $H_A=H_B=0$: i.e., states which do not violate the minimax constraints.

Now, given the states where $H_A=H_B=0$, we now want the ground state of $H_A+H_B+H_C$ to be the state where the fewest number of edges are colored.   To do this, we simply let \begin{equation}
H_C = C\sum_e x_e
\end{equation}count the number of colored edges.    Here $C$ is an energy scale chosen to be small enough such that it is never energetically favorable to violate the constraints imposed by either the $H_A$ or $H_B$ terms:  one requires $C<B$, since there is an energy penalty of $B$ associated to each edge which could be colored, but isn't.   The term with the smallest $H_C$ has the smallest number of edges, and is clearly the solution to the minimax problem.   Each ground state of this spin model is equivalent to a solution of the minimax problem.

\section{Problems with Inequalities}\label{sec4}
We now turn to NP problems whose formulations as Ising models are more subtle, due to the fact that constraints involve inequalities as opposed to equalities.    These constraints can be re-written as constraints only involving equalities by an expansion of the number of spins.

As with partitioning problems, we will find that these Hamiltonians require embedding highly connected graphs onto a quantum device. This may limit their usability on current hardware.
\subsection{Set Cover}
Consider a set $U=\lbrace 1,\ldots, n\rbrace$, with sets $V_i \subseteq U$ ($i=1,\ldots, N$) such that \begin{equation}
U = \bigcup_{i=1}^N V_\alpha.
\end{equation}The set covering problem is to find the smallest possible number of $V_i$s, such that the union of them is equal to $U$.    This is a generalization of the exact covering problem, where we do not care if some $\alpha \in U$ shows up in multiple sets  $V_i$; finding the smallest number of sets which ``cover" $U$ is NP-hard \cite{karp}.

Let us denote $x_i$ to be a binary  variable which is 1 if set $i$ is included, and 0 if set $i$ is not included.    Let us then denote $x_{\alpha,m}$ to be a binary variable which is 1 if the number of $V_i$s which include element $\alpha$ is $m\ge 1$, and 0 otherwise.    Set $H=H_A+H_B$.   Our first energy imposes the constraints that exactly one $x_{\alpha,m}$ must be 1, since each element of $U$ must be included a fixed number of times, and that the number of times that we claimed $\alpha$ was included is in fact equal to the number of $V_i$ we have included, with $\alpha$ as an element: \begin{equation}
H_A = A\sum_{\alpha=1}^n \left(1-\sum_{m=1}^N x_{\alpha,m}\right)^2+A\sum_{\alpha=1}^n \left(\sum_{m=1}^N mx_{\alpha,m}-\sum_{i: \alpha \in V_i} x_i \right)^2.
\end{equation}
Finally, we minimize over the number of $V_\alpha$s included: \begin{equation}
H_B =B\sum_{i=1}^N x_i,
\end{equation}with $0<B < A$ required to never violate the constraints of $H_A$ (the worst case is that one set must be included to obtain one element of $U$; the change in $H$ if we include this last element is $B-A$, which must be negative).

Let $M\le N$ be the maximal number of sets which contain any given element of $U$; then $N$ $x_i$s are required, and $n\lfloor 1+ \log M \rfloor$ spins are required (using the trick described earlier) for the $x_{\alpha,m}$ spins; the total number is therefore $N+n\lfloor 1+\log M\rfloor$ spins.

\subsection{Knapsack with Integer Weights}
The knapsack problem is the following problem:  we have a list of $N$ objects, labeled by indices $\alpha$, with the weight of each object given by $w_\alpha$, and its value given by $c_\alpha$, and we have a knapsack which can only carry weight $W$.   If $x_\alpha$ is a binary variable denoting whether (1) or not (0) object $\alpha$ is contained in the knapsack, the total weight in the knapsack is \begin{equation}
\mathcal{W} = \sum_{\alpha=1}^N w_\alpha x_\alpha
\end{equation}and the total cost is \begin{equation}
\mathcal{C} = \sum_{\alpha=1}^N c_\alpha x_\alpha.
\end{equation}The NP-hard \cite{karp} knapsack problem asks us to maximize $\mathcal{C}$ subject to the constraint that $\mathcal{W} \le W$.   It has a huge variety of applications, particularly in economics and finance \cite{Kellerer2005}.

Let $y_n$ for $1\le n\le W$ denote a binary variable which is 1 if the final weight of the knapsack is $n$, and 0 otherwise.   Our solution consists of letting $H=H_A+H_B$, with \begin{equation}
H_A = A\left(1-\sum_{n=1}^W y_n\right)^2+A\left(\sum_{n=1}^W ny_n - \sum_\alpha w_\alpha x_\alpha \right)^2
\end{equation}which enforces that the weight can only take on one value and that the weight of the objects in the knapsack equals the value we claimed it did, and finally \begin{equation}
H_B = -B\sum_\alpha c_\alpha x_\alpha.
\end{equation}As we require that it is not possible to find a solution where $H_A$ is weakly violated at the expense of $H_B$ becoming more negative, we require $0<B\max(c_\alpha) < A$ (adding one item to the knapsack, which makes it too heavy, is not allowed).   The number of spins required is (using the log trick) $N + \lfloor 1 + \log W\rfloor$.

\section{Coloring Problems}\label{sec5}
We now turn to coloring problems.   Naively, coloring problems are often best phrased as Potts models \cite{fywu}, where the spins can take on more than two values, but these classical Potts models can be converted to classical Ising models with an expansion of the number of spins.   This simple trick forms the basis for our solutions to this class of problems.

\subsection{Graph Coloring}
Given an undirected graph $G=(V,E)$, and a set of $n$ colors, is it possible to color each vertex in the graph with a specific color, such that no edge connects two vertices of the same color?   This is one of the more famous NP-complete \cite{karp} problems, as one can think of it as the generalization of  the problem of how many colors are needed to color a map, such that no two countries which share a border have  the same color.   Of course, in this special case,\footnote{The graphs are \emph{planar} -- the vertices can be realized by points on $\mathbb{R}^2$, and the edges as line segments between them, such that no two line segments intersect (except at a vertex).} one can prove that there is always a coloring for $n\ge 4$ \cite{appel1, appel2}.   This  problem is called the graph coloring problem.

Our solution consists of the following:   we denote $x_{v,i}$ to be a binary variable which is 1 if vertex $v$ is colored with color $i$, and 0 otherwise.     The energy is\begin{equation}
H = A\sum_v \left(1-\sum_{i=1}^n x_{v,i}\right)^2 + A\sum_{(uv)\in E} \sum_{i=1}^n x_{u,i}x_{v,i}.
\end{equation}The first term enforces the constraint that each vertex has exactly one color, and provides an energy penalty each time this is violated, and the second term gives an energy penalty each time an edge connects two vertices of the same color.   If there is a ground state of this model with $H=0$, then there is a solution to the coloring problem on this graph with $n$ colors.   We can also read off the color of each node (in one such coloring scheme) by looking at which $x$s are 1.   Note that the number of spins can be slightly reduced, since there is a permutation symmetry among colorings, by choosing a specific node in the graph to have the color 1, and one of its neighbors to have the color 2, for example.    The total number of spins required is thus $nN$.  

\subsection{Clique Cover}
The clique cover problem, for an undirected graph $G=(V,E)$, is the following: given $n$ colors, we assign a distinct color to each vertex of the graph.   Let $W_1,\ldots, W_n$ be the subsets of $V$ corresponding to each color, and $E_{W_1},\ldots, E_{W_n}$ the edge set restricted to edges between vertices in the $W_i$ sets.   The clique cover problem asks whether or not $(W_i, E_{W_i})$ is a complete graph for each $W_i$ (i.e., does each set of colored vertices form a clique?).   This problem is known to be NP-complete \cite{karp}.

Our solution is very similar to the graph coloring problem.   Again, we employ the same binary variables as for graph coloring, and use a Hamiltonian very similar to the cliques problem: \begin{equation}
H=A\sum_v \left(1-\sum_{i=1}^n x_{v,i}\right)^2+B\sum_{i=1}^n  \left[\frac{1}{2}\left(-1+\sum_v x_{v,i}\right)\sum_v x_{v,i} - \sum_{(uv)\in E} x_{u,i}x_{v,i} \right].
\end{equation}   The first term enforces the constraint that each vertex has exactly one color by giving an energy penalty each time this constraint is violated.   In the second term, since the sum over $v$ of  $x_{v,i}$ counts the number of nodes with color $i$, the first sum counts highest possible number of edges that could exist with color $i$.   The second term then checks if, in fact, this number of edges does in fact exist.  Thus $H=0$ if and only if the clique cover problem is solved by the given coloring.    If a ground state exists with $H=0$, there is a solution to the clique covering problem.   The discussion on the required ratio $A/B$ to encode the correct solution is analogous to the discussion for the cliques problem.   The total number of spins required is $nN$.

\subsection{Job Sequencing with Integer Lengths}
The job sequencing problem is as follows:   we are given a list of $N$ jobs for $m$ computer clusters.   Each job $i$ has length $L_i$.   How can each job be assigned to a computer in the cluster such that, if the set of jobs on cluster $\alpha$ is $V_\alpha$, then the length of that cluster, defined as\begin{equation}
M_\alpha \equiv \sum_{i\in V_\alpha} L_i,
\end{equation}are chosen such that $\max(M_\alpha)$ is minimized?  Essentially, this means that if we run all of the jobs simultaneously, all jobs will have finished running in the shortest amount of time.   This is NP-hard \cite{karp}, and there is a decision version (is $\max(M_\alpha) \le M_0$?) which is NP-complete.  We assume that $L_i \in \mathbb{N}$.   

To do this, we will begin by demanding that without loss of generality, $M_1 \ge M_\alpha$ for any $\alpha$.   Introduce the variables $x_{i,\alpha}$ which are 1 if job $i$ is added to computer $\alpha$, and 0 otherwise, and the variables $y_{n,\alpha}$ for $\alpha \ne 1$ and $n\ge 0$, which is 1 if the difference $M_1-M_\alpha = n$.    Then the Hamiltonian \begin{equation}
H_A = A\sum_{i=1}^N \left(1-\sum_\alpha x_{i,\alpha}\right)^2 + A\sum_{\alpha=1}^m \left(\sum_{n=1}^{\mathcal{M}} ny_{n,\alpha}  + \sum_i L_i (x_{i,\alpha}-x_{i,1})\right)^2
\end{equation}encodes that each job can be given to exactly one computer, and that no computer can have a longer total length than computer 1.   The number $\mathcal{M}$ must be chosen by the user, and is related to the number of auxiliary spins required to adequately impose the length constraints that $M_1\ge M_\alpha$:  in the worst case, it is given by $N\max(L_i)$.  To find the minimal maximal length $M_1$, we just use \begin{equation}
H_B =B\sum_i L_i x_{i,1}.
\end{equation}Similarly to finding bounds on $A/B$ for the knapsack problem, for this Hamiltonian to encode the solution to the problem, we require (in the worst case) $0<B\max(L_i)<A$.    Using the log trick, the number of spins required here is $mN + (m-1)
\lfloor 1 + \log \mathcal{M}\rfloor $.

\section{Hamiltonian Cycles}\label{sec6}
In this section, we describe the solution to the (undirected or directed) Hamiltonian cycles problem, and subsequently the traveling salesman problem, which for the Ising spin glass formulation, is a trivial extension.
\subsection{Hamiltonian Cycles and Paths}
Let $G=(V,E)$, and $N=|V|$.   The graph can either be directed or undirected; our method of  solution will not change.    The Hamiltonian path problem is as follows:   starting at some node in the graph, can one travel along an edge, visiting other nodes in the graph, such that one can reach every single node in the graph without ever returning to the same node twice?   The Hamiltonian cycles problem asks that, in addition, the traveler can return to the starting point from the last node he visits.   Hamiltonian cycles is a generalization of the famous K\"onigsberg bridge problem \cite{mezard}, and is NP-complete \cite{karp}.

Without loss of generality, let us label the vertices $1,\ldots, N$, and take the edge set $(uv)$ to be directed -- i.e., the order $uv$ matters.   It is trivial to extend to undirected graphs, by just considering a directed graph with $(vu)$ added to the edge set whenever $(uv)$ is added to the edge set.   Our solution will use $N^2$ bits $x_{v,i}$, where $v$ represents the vertex and $i$ represents its order in a prospective cycle.    Our energy will have three components.   The first two things we require are that every vertex can only appear once in a cycle, and that there must be a $j^{\mathrm{th}}$ node in the cycle for each $j$.   Finally, for the nodes in our prospective ordering,  if $x_{u,j}$ and $x_{v,j+1}$ are both 1, then there should be an energy penalty if $(uv)\notin E$.    Note that $N+1$ should be read as 1, in the expressions below, if we are solving the cycles problem.   These are encoded in the Hamiltonian: \begin{equation}
H = A\sum_{v=1}^n\left(1-\sum_{j=1}^N x_{v,j}\right)^2 + A\sum_{j=1}^n \left(1-\sum_{v=1}^N x_{v,j}\right)^2 + A\sum_{(uv)\notin E} \sum_{j=1}^N x_{u,j}x_{v,j+1}.
\end{equation}
$A>0$ is a constant.   It is clear that a ground state of this system has $H=0$ only if we have an ordering of vertices where each vertex is only included once, and adjacent vertices in the cycle have edges on the graph -- i.e., we have a Hamiltonian cycle.

To solve the Hamiltonian path problem instead, restrict the last sum over $j$ above from 1 to $N-1$; we do not care about whether or not the first and last nodes are also connected.   $N^2$ spins are requied to solve this problem.

It is straightforward to slightly reduce the size of the state space for the Hamiltonian cycles problem as follows:  it is clear that node 1 must always be included in a Hamiltonian cycle, and without loss of generality we can set $x_{1,i}=\delta_{1,i}$:  this just means that the overall ordering of the cycle is chosen so that node 1 comes first.   This reduces the number of spins to $(N-1)^2$.
\subsection{Traveling Salesman}
The traveling salesman problem for a graph $G=(V,E)$, where each edge $uv$ in the graph has a weight $W_{uv}$ associated to it, is to find the Hamiltonian cycle such that the sum of the weights of each edge in the cycle is minimized.  Typically, the traveling salesman problem assumes a complete graph, but we have the technology developed to solve it on a more arbitrary graph.    The decision problem (does a path of total weight $\le W$ exist?) is NP-complete \cite{karp}.

To solve this problem, we use $H=H_A+H_B$, with $H_A$ the Hamiltonian given for the directed (or undirected) Hamiltonian cycles problem.   We then simply add \begin{equation}
H_B = B\sum_{(uv)\in E} W_{uv}\sum_{j=1}^N x_{u,j}x_{v,j+1}.
\end{equation} with $B$ small enough that it is never favorable to violate the constraints of $H_A$; one such constraint is $0<B \max(W_{uv}) < A$ (we assume in complete generality $W_{uv}\ge 0$ for each $(uv)\in E$).\footnote{One can also encode graph structure by assuming a complete graph (this allows one to neglect the third term in $H_A$), but choosing the weights of the ``non-existent" edges to obey $W_{uv\notin E} \ge N \max(W_{uv\in E})$.  As $W_{uv}$ is not defined if $(uv)\notin E$, these are in fact two identical interpretations.}   If the traveling salesman does not have to return to his starting position, we can restrict the sum over $j$ from $1$ to $N-1$, as before.   As with Hamiltonian cycles, $(N-1)^2$ spins are required, as we may fix node 1 to appear first in the cycle.

\section{Tree Problems}\label{sec7}
The most subtle NP problems to solve with Ising models are problems which require finding connected tree subgraphs of larger graphs.\footnote{ A tree is a graph with no cycles.  A cycle is set of vertices $v_1,\ldots, v_n$ with $(v_1v_2),\ldots, (v_{n-1}v_n), (v_nv_1) \in E$.  It is easy to check that if $(V,E)$ is a tree, $|E|=|V|-1$.}   Because determining whether a subgraph is a tree requires global information about the connectivity of a graph, we will rely on similar tricks to what we used to write down Hamiltonian cycles as an Ising model.  
\subsection{Minimal Spanning Tree with a Maximal Degree Constraint}
The minimal spanning tree problem is the following:   given an undirected graph $G=(V,E)$, where each edge $(uv) \in E$ is associated  with a cost $c_{uv}$, what is the tree $T\subseteq G$, which contains all vertices, such that the cost of $T$, defined as \begin{equation}
c(T) \equiv \sum_{(uv)\in E_T} c_{uv},
\end{equation}is minimized (if such a tree exists)?   Without loss of generality, we take $c_{uv}>0$ in this subsection (a positive constant can always be added to each $c_{uv}$ ensure that the smallest value of $c_{uv}$ is strictly positive, without changing the trees $T$ which solve the problem).   We will also add a degree constraint, that each degree in $T$ be $\le \Delta$.   This makes the problem NP-hard, with a corresponding NP-complete decision problem \cite{karp}.

To solve this problem, we place a binary variable $y_e$ on each edge to determine whether or not that edge is included in $T$: \begin{equation}
y_e \equiv \left\lbrace\begin{array}{ll} 1 &\ e\in E_T \\ 0 &\ \text{otherwise} \end{array}\right..
\end{equation}We also place a large number of binary variables $x_{v,i}$ on each vertex, and $x_{uv,i}, x_{vu,i}$ on edge $(uv)$ (these are distinct spins):   the number $i=0,1,\ldots, N/2$ will be used to keep track of the depth a node in the tree, and if $x_{uv}=1$, it means that $u$ is closer to the root than $v$, and if $x_{vu}=1$ it means that $v$ is closer to the root.   Finally, we use another variable $z_{v,i}$ ($i=1,\ldots \Delta$) to count the number of  degrees of each node.      We now use energy $H=H_A+H_B$, where the terms in $H_A$ are used to impose the constraints that:  there is exactly one root to the tree, each vertex has a depth, each bond has a depth, and its two vertices must be at different heights, the tree is connected (i.e., exactly one edge to a non-root vertex comes from a vertex at lower depth), each node can have at most $\Delta$ edges, and each edge at depth $i$ points between a node at depth $i-1$ and $i$, respectively:   \begin{align}
H_A&= A\left(1-\sum_v x_{v,0}\right)^2 + A\sum_v \left(1-\sum_i x_{v,i}\right)^2 + A\sum_{uv\in E} \left(y_{uv}-\sum_i (x_{uv,i}+x_{vu,i})\right)^2  \notag \\
&\;\;\;\;\; + A\sum_v \sum_{i=1}^{N/2}\left(x_{v,i}-\sum_{u:(uv)\in E} x_{uv,i}\right)^2 + A \sum_v \left(\sum_{j=1}^\Delta jz_{v,j}-\sum_{u:(uv)\in E}\sum_i (x_{uv,i}+x_{vu,i})\right)^2   \notag \\
&\;\;\;\;\; +  A\sum_{(uv),(vu)\in E} \sum_{i=1}^{N/2} x_{uv,i}(2-x_{u,i-1}-x_{v,i})
\end{align}
The ground states with $H_A=0$ are trees which include every vertex.   In the last term in the sum, remember that $x_{uv,i}$ and $x_{vu,i}$ are both spins that are included for each edge; the notation in the summation is meant to remind us of this.   We then add \begin{equation}
H_B =B\sum_{uv,vu\in E}\sum_{i=1}^{N/2} c_{uv} x_{uv,i}.
\end{equation}In order to solve the correct problem, we need to make sure that we never remove any $x_{uv,i}$ from $H_B$ in order to have a more negative total $H$.   As each constraint in $H_A$ contributes an energy $\ge A$ if it is violated, we conclude that setting $0<B\max(c_{uv}) < A$ is sufficient.   The minimum of $E$ will find the minimal spanning tree, subject to the degree constraint.   

The number of spins required is  $|V|(\lfloor |V| +1\rfloor+2)/2 + |E|(|V|+1) + |V|\lfloor 1 + \log \Delta \rfloor$.  The maximal possible number of edges on any graph is $|E| = \mathrm{O}(|V|^2)$, so this Ising formulation may require a cubic number of spins in the size of the vertex set.

\subsection{Steiner Trees}
The NP-hard \cite{karp} Steiner tree problem is somewhat similar to the problem above:  given our costs $c_{uv}$, we want to find a minimal spanning tree for a subset $U\subset V$ of the vertices (i.e., a tree such that the sum of $c_{uv}$s along all included edges is minimal).   We no longer impose degree constraints; the problem turns out to be ``hard" already, as we now allow for the possibility of not including nodes which are not in $U$.

To solve this by finding the ground state of an Ising model, we use the same Hamiltonian as for the minimal spanning tree, except we add binary variables $y_v$ for $v\notin U$ which determine whether or not a node $v$ is included in the tree.   We use the Hamiltonian $H=H_A+H_B$, where $H_A$ enforces constraints similarly to in the previous case: \begin{align}
H_A&= A\left(1-\sum_v x_{v,0}\right)^2   + A\sum_{v\in U} \left(1-\sum_i x_{v,i}\right)^2 + A\sum_{v\notin U} \left(y_v-\sum_i x_{v,i}\right)^2  \notag \\
&\;\;\;\;\; + A\sum_v \sum_{i=1}^{N/2}\left(x_{v,i}-\sum_{(uv)\in E} x_{uv,i}\right)^2 +  A\sum_{uv,vu\in E} \sum_{i=1}^{N/2} x_{uv,i}(2-x_{u,i-1}-x_{v,i}) \notag \\
&\;\;\;\;\; + A\sum_{uv\in E} \left(y_{uv}-\sum_i (x_{uv,i}+x_{vu,i})\right)^2
\end{align}We then use $H_B$ from the previous model to determine the minimum weight tree; the same constraints on $A/B$ apply.   The number of spins is $|V|(\lfloor |V| +1\rfloor+4+2|E|)/2 + |E|$.

\subsection{Directed Feedback Vertex Set} 
A feedback vertex set for a directed graph $G=(V,E)$ is a subset $F\subset V$ such that the subgraph $(V-F, \partial(V-F))$ is acyclic (has no cycles).   We will refer to $F$ as the feedback set.   Solving a decision problem for whether or not a feedback set exists for $|F|\le k$ is NP-complete \cite{karp}.   We solve the optimization problem of finding the smallest size of the feedback set first for a directed graph -- the extension to an undirected graph will be a bit more involved.

Before solving this problem, it will help to prove two lemmas.   The first lemma is quite simple:  there exists a node in a directed acyclic graph which is not the end point of any edges.   Suppose that for each vertex, there was an edge that ends on that vertex.  Then pick an arbitrary vertex, pick any edge ending on that vertex, and follow that edge in reverse to the starting vertex.   Repeat this process more than $N$ times, and a simple counting argument implies that we must have visited the same node more than once, at least once.   Thus, we have traversed a cycle in reverse, which contradicts our assumption.

The second lemma is as follows:  a directed graph $G=(V,E)$ is acyclic if and only if there is a height function $h:V\rightarrow \mathbb{N}$ such that if $uv\in E$, $h(u)<h(v)$:  i.e., every edge points from a node at lower height to one at higher height.      That height function existence implies acyclic is easiest to prove using the contrapositive:  suppose that a graph is cyclic.   Then on a cycle of edges, we have \begin{equation}
0 < \sum [h(u_{i+1}) - h(u_i)]  = h(u_1) - h(u_n) + h(u_n) - h(u_{n-1}) + \cdots -h(u_1)=0
\end{equation}is a contradiction.     To prove that an acyclic graph has a height function, we construct one recursively.   Using our first lemma, we know that there exists a vertex $u$ with only outgoing edges, so let us call $h(u)=1$.   For any other vertex, we will call the height of that vertex $h(v) = 1+h^\prime(v)$, where $h^\prime(v)$ is found by repeating this process on the graph with node $u$ removed (which must also be acyclic).   It is clear this process will terminate and assign exactly one node height $i$ for each integer $1\le i\le |V|$.

We can now exploit this lemma to write down an Ising spin formulation of this problem.   We place a binary variable $y_v$ on each vertex, which is 0 if $v$ is part of the feedback set, and 1 otherwise.   We then place a binary variable $x_{v,i}$ on each vertex, which is 1 if vertex $v$ is at height $i$.    So far the heights $i$ are arbitrary, and the requirement that a height function be valid will be imposed by the energy.    The energy functional we use is  $H=H_A+H_B$ where\begin{equation}
H_A = A\sum_v \left(y_v - \sum_i x_{v,i}\right)^2 + A \sum_{uv\in E} \sum_{i\ge j} x_{u,i}x_{v,j}.
\end{equation}
The first term ensures that if a vertex is not part of the feedback set, it has a well-defined height; the second term ensures that an edge only connects a node with lower height to a node at higher height.    We then find the smallest possible feedback set by adding \begin{equation}
H_B = B\sum_v (1-y_v).
\end{equation}
In order to solve the correct problem, we cannot add too few nodes to the feedback set.  If we set $y_v=1$ for a node which should be part of the feedback set, we find an energy penalty of $A$ from $H_A$, and a gain of $B$ from $H_B$.   We conclude that $B<A$ is sufficient to ensure we solve the correct problem.   We see that $|V|(|V|+1)$ spins are required.

\subsection{Undirected Feedback Vertex Set}
The extension to undirected graphs requires a bit more care.   In this case, we have to be careful because there is no a priori distinction on whether the height of one end of an edge is smaller or larger than the other -- this makes the problem much more involved, at first sight.   Furthermore, it is not true that a directed acyclic graph is acyclic if the orientation of edges is ignored.  However, for an undirected graph, we also know that a feedback vertex set must reduce the graph to trees, although there is no longer a requirement that these trees are connected (this is called a forest).   With this in mind, we find that the problem is actually extremely similar to minimal spanning tree, but without degree constraints or connectivity constraints.     The new subtlety, however, is that we cannot remove edges.

To solve this problem, we do the following:  introduce a binary variable $x_{v,i}$, which is 1 if $v$ is a vertex in any tree (anywhere in the forest) at depth $i$, and 0 otherwise.   However, to account for the fact that we may remove vertices, we will allow for $y_{v}=1$ if $v$ is part of the feedback vertex set, and 0 otherwise.   We do a similar thing for edges:  we consider $x_{uv,i}, x_{vu,i}$ to be defined as before when $i>0$.   We also define the variables $y_{uv}, y_{vu}$, which we take to be 1 when the ending node of the ``directed" edge is in the feedback vertex set.   Now, we can write down a very similar energy to the minimal spanning tree:  \begin{align}
H_A &= A\sum_v \left(1-y_v-\sum_i x_{v,i}\right)^2 + A\sum_{uv\in E} \left(1-\sum_i (x_{uv,i} + x_{vu,i}+y_{uv}+y_{vu})\right)^2 +A \sum_{uv \in E} (y_{uv}-y_{v})^2 \notag \\
&\;\;\;\;\; + A\sum_v\sum_{i>0}\left(x_{v,i} - \sum_{u:uv\in E}x_{uv,i}\right)^2 + A\sum_{uv,vu\in E} \sum_{i>0} x_{uv,i}(2-x_{u,i-1}-x_{v,i})
\end{align}
The changes are as follows:  we no longer constrain only 1 node to be the root, or constrain the degree of a vertex -- however, we have to add a new term to ensure that edges are only ignored in the tree constraint if they point to a node in the feedback set.    We then add \begin{equation}
H_B = B\sum_v y_{v}
\end{equation}with $B<A$ required so that the $A$ constraints are never violated.   This counts the number of nodes in the feedback set, so thus $H$ is minimized when $H_B$ is smallest -- i.e., we have to remove the fewest number of nodes.   The number of spins required is $(|E|+|V|)\lceil (|V|+3)/2 \rceil $.\footnote{This follows from the fact that the sum over $i$s is $\lceil (|V|+1)/2\rceil$; then we account for the $y$s.}

The recent paper \cite{zhou} has a more efficient implementation of a mapping, for use in understanding random ensembles of this problem by the replica method.   Unfortuntaely, this technique is not efficient for AQO; the Hamiltonian contains $N$-body terms.

\subsection{Feedback Edge Set}
For a directed graph, the feedback edge set problem is to find the smallest set of edges $F\subset E$ such that $(V,E-F)$ is a directed acyclic graph.    It is known to be NP-hard \cite{karp}.\footnote{It is in P if the graph is undirected however.} Our solution will be somewhat similar to the directed feedback vertex set.    We place a binary variable $y_{uv}$ on each edge, which is 1 if $uv\notin F$, and define $x_{uv,i}$ to be 1  if both $y_{uv}=1$ and the height of node $u$ is $i$.    We also add a binary variable $x_{v,i}$, as for the feedback vertex set.   Our constraint energy must then enforce that:  each vertex and included edge has a well-defined height, and that each edge points from a lower height to a higher height: \begin{equation}
H_A = A\sum_v \left(1-\sum_i x_{v,i}\right)^2 + A\sum_{uv\in E}\left(y_{uv}-\sum_i x_{uv,i}\right)^2 + A\sum_{uv}\sum_i x_{uv,i}\left(2-x_{u,i}-\sum_{j>i} x_{v,j}\right).
\end{equation}
We then use \begin{equation}
H_B = B\sum_{uv\in E} (1-y_{uv})
\end{equation}to count the number of edges in $F$ -- it is minimized when this number is smallest.   As before, one needs $B<A$ to encode the correct problem.   The number of spins required is $|E| + |V|(|V|+|E|)$.

\section{Graph Isomorphisms}\label{sec8}
Graphs $G_1$ and $G_2$, with $N$ vertices each, are isomorphic if there is a labeling of vertices $1,\ldots, N$ in each graph such that the adjacency matrices for the graphs is identical.    More carefully:   any graph $G=(V,E)$, with vertices labeled as $1,\ldots , N$, has an $N\times N$ adjacency matrix $\mathsf{A}$ with \begin{equation}
A_{ij} = \left\lbrace\begin{array}{ll} 1 &\ (ij)\in E, \\ 0 &\ (ij)\notin E. \end{array}\right.,
\end{equation}
which contains all information about the edge set $E$.   Let $\mathsf{A}_{1,2}$ be the adjacency matrices of graphs $G_{1,2}$.   If there is a permutation matrix $\mathsf{P}$ such that $\mathsf{A}_2 = \mathsf{P}^{\mathsf{T}} \mathsf{A}_1 \mathsf{P}$, then we say $G_{1,2}$ are isomorphic.

The question of whether two graphs $G_1=(V_1,E_1)$ and $G_2=(V_2,E_2)$ are isomorphic is believed to be hard, but its classification into a complexity class is still a mystery \cite{johnson}.   Since it is (in practice) a hard problem, let us nonetheless describe an Ising formulation for it.   An isomorphism is only possible if $|V_1|=|V_2|\equiv N$, so we will restrict ourselves to this case, and without loss of generality, we label the vertices of $G_1$ with $1,\ldots, N$.    

We write this as an Ising model as follows.   Let us describe a proposed isomorphism through binary variables $x_{v,i}$ which is 1 if vertex $v$ in $G_2$ gets mapped to vertex $i$ in $G_1$.   The energy \begin{equation}
H_A = A\sum_v \left(1-\sum_i x_{v,i}\right)^2 + A\sum_i \left(1-\sum_v x_{v,i}\right)^2
\end{equation}ensures that this map is bijective.   We then use an energy \begin{equation}
H_B =  B\sum_{ij\notin E_1} \sum_{uv\in E_2} x_{u,i}x_{v,j} +B\sum_{ij\in E_1} \sum_{uv\notin E_2} x_{u,i}x_{v,j} 
\end{equation} to penalize a bad mapping:  i.e. an edge that is not in $G_1$ is in $G_2$, or an edge that is in $G_1$ is not in $G_2$.   As usual, assume $A,B>0$.   If the ground state of this Hamiltonian has $H=0$, there is an isomorphism.     $N^2$ spins are required.

An approximate algorithm that uses quantum annealing to distinguish between non-isomorphic graphs via the spectra of graph-dependent Hamiltonians was presented in \cite{hen2}.

\section{Conclusions}

 The focus of research into AQO has essentially been on NP-complete/hard problems, because the Ising model is NP-hard, and because computer scientists have struggled to find efficient ways of solving these problems.    In this paper, we have presented strategies for mapping a wide variety of NP problems to Ising spin glasses, exemplified by a demonstration of a glass for each of Karp's 21 NP-complete problems.    It is an open question the extent to which AQO will help provide efficient solutions for these problems, whether these solutions are exact or approximate.
 
 However, physicists are interested in building a universal quantum computer which is capable of solving much more than just Ising models.   As an example, a universal quantum computer would also reduce the time for searching an unsorted list of $N$ items from $\mathrm{O}(N)$ to $\mathrm{O}(\sqrt{N})$ \cite{Nielsen2000}.   This would be incredibly useful for many practical applications, despite the fact that searching is an easy linear time algorithm.   Analogously, it may be the case that there exists a family of ``easy" problems which AQO can solve in polynomial time, yet more efficiently than a classical polynomial time algorithm.   This statement may even be true with Ising-implementing AQO hardware, although if so it is not obvious.
 
It is certainly the case that an AQO-implementing device can be used to solve easy problems.   Consider the simple problem of finding the largest integer in a list $n_1,\ldots,n_N$ (this is the searching algorithm that a universal quantum computer can perform efficiently).   Introducing binary variables $x_i$ for $i=1,\ldots, N$, the Ising model \begin{equation}
H = A\left(1-\sum_i x_i\right)^2 - B\sum_i n_ix_i
\end{equation}for $A>B\max(n_i)$ solves this problem.   In fact, this problem looks somewhat like an instance of the random field Ising model on a complete graph, and yet this has a very simple $\mathrm{O}(N)$ classical algorithm.   It would surely take longer to program this algorithm into a quantum device than to solve the problem itself.   

The above example demonstrates that sometimes the ``hardness" of a problem can be deceptive -- one can phrase something that is easy in a way which makes it seem hard.   It is worth discussing more closely the hardness of NP problems, because it turns out that sometimes, NP problems can be easier than they first appear.   To be NP-complete but not P (if $\mathrm{P}\ne \mathrm{NP}$) one only needs a small family of instances of the problem to be unsolvable in polynomial time by a deterministic algorithm.  However, typical instances may not be so hard.   Many popular NP problems can almost surely be solved exactly in polynomial time on large random  instances \cite{Beier2003, krivelevich},\footnote{One has to be careful with the phrases ``random" and ``typical", as this immediately implies a probability distribution over a space of problem instances.  This probability distribution may place vanishingly small probability on a set of relevant instances for any given application.  For the simple probability distributions used in these papers, it is highly non-trivial that most instances turn out to be solvable in polynomial time.} and there exist randomized algorithms for some NP problems which can get arbitrarily close to a solution with arbitrarily low failure probability in polynomial time \cite{dyer, vazirani} (though multiplicative coefficients or polynomial exponents must diverge as the failure probability and/or error on determining the ground state tends to zero, if $\mathrm{P}\ne \mathrm{NP}$).   In addition, popular algorithms in P, like matrix decomposition, may serve as the bottlenecks of \emph{practical} computations, and should not be thought of as ``easy".    Typical instances approach the asymptotic bounds on worst-case runtimes, in contrast to the case for some NP problems; many recent developments focus on randomized algorithms  \cite{rokpnas, martinsson, lucas}.

The Hamiltonians of this paper may be deceptively ``hard" -- this can mean that they involve too many spins.    Another possibility is that these Hamiltonians have small spectral gaps, and that alternative choices have much larger spectral gaps -- this is a question we have not addressed at all in this paper.     Studying how to simplify quantum algorithms, and more importantly increase energy gaps (and thus reduce $T$), even by constant factors, is a much needed endeavor.
\section*{Acknowledgements}
A.L. is supported by the Smith Family Graduate Science and Engineering Fellowship at Harvard.    

He would like to thank Robert Lucas for pointing out that a compendium of ways to map famous NP problems to Ising glasses was lacking, Jacob Biamonte for encouraging publication, and Vicky Choi, Jacob Sanders, Federico Spedalieri, John Tran, and especially the reviewers, for many helpful comments on AQO and computer science.

\bibliographystyle{unsrt}
\addcontentsline{toc}{section}{References}
\bibliography{isingbib}

\end{document}